\documentclass[12pt]{article}
\usepackage{amsmath}
\usepackage{graphicx,psfrag,epsf}
\usepackage{enumerate}
\usepackage{natbib}
\usepackage{textcomp}
\usepackage[hyphens]{url} 
\usepackage{hyperref}

\newcommand{\blind}{0}

\begin{document}

\def\spacingset#1{\renewcommand{\baselinestretch}%
{#1}\small\normalsize} \spacingset{1}


\if0\blind
{
  \title{\bf Bringing Visual Inference to the Classroom}

  \author{
        Adam Loy \\
    Department of Mathematics and Statistics, Carleton College\\
      }
  \maketitle
} \fi

\if1\blind
{
  \bigskip
  \bigskip
  \bigskip
  \begin{center}
    {\LARGE\bf Bringing Visual Inference to the Classroom}
  \end{center}
  \medskip
} \fi

\bigskip
\begin{abstract}
In the classroom, we traditionally visualize inferential concepts using
static graphics or interactive apps. For example, there is a long
history of using apps to visualize sampling distributions. Recent
developments in statistical graphics have created an opportunity to
bring additional visualizations into the classroom to hone student
understanding. Specifically, the lineup protocol for visual inference
provides a framework for students see the difference between signal and
noise by embedding a plot of observed data in a field of null (noise)
plots. Lineups have proven valuable in visualizing
randomization/permutation tests, diagnosing models, and even conducting
valid inference when distributional assumptions break down. This paper
provides an overview of how the lineup protocol for visual inference can
be used to hone understanding of key statistical topics throughout the
statistics curricula.
\end{abstract}

\noindent%
\vfill

\newpage
\spacingset{1.45} 

\hypertarget{introduction}{%
\section{Introduction}\label{introduction}}

Recent years have seen a great deal of innovation in how we teach
statistics as we strive to overcome what \citet{Cobb-2007uo} termed
``the tyranny of the computable.'' Most notably, simulation-based
pedagogies for the first course have been proposed and validated
\citep{Cobb-2007uo, Tintle-2011vo, Tintle-2012td, Maurer-2014te, Tintle2014-vt, Hildreth2018}.
These simulation-based pedagogies have also been used in mathematical
statistics \citep{chihara2011, Cobb2011-vz} and \citet{Tintle2015-yv}
argue that they should be used throughout the entire curriculum.

In addition to changes to how we introduce inference, there have also
been changes to the computational toolkit we use throughout the
statistic curricula. At the introductory level, numerous toolkits are
commonplace depending on the objectives and audience of the course. Web
apps are commonly used when students may not have access to their own
computers, or simply to lower the technical barriers to entry. Examples
include StatKey \citep{Lock2017}, the \emph{Introduction to Statistical
Investigations} applets \citep{tintle2015}, and the shiny apps from
\citet{agresti2017}. These apps allow students to explore course
concepts without getting into the computational weeds. For courses
exploring both the concepts and implementation in a realistic
data-analytic workflow, R \citep{r} is a common open-source choice and
multiple R packages have been developed to lower the barriers to entry
for students, notably the mosaic \citep{Pruim2017-uc}, ggformula
\citep{ggformula}, and infer \citep{infer}.

The above developments enabled the statistics education community to
address key recommendations made in the GAISE report \citep{gaise2016}.
The simulation-based curriculum has focused attention on teaching
statistical thinking and fostering conceptual understanding before
delving into the mathematical details. An improved computational toolkit
has enabled students to use technology to explore concepts, such as a
sampling and permutation distributions, and to analyze data.

While the use of the simulation-based curriculum has helped students
focus on the underlying ideas of statistical inference, little has
changed about the way we help students \emph{visualize} inference.
Specifically, we still have students grapple with null/reference
distributions in hypothesis testing and sampling distributions for
estimation while they are trying to hone their intuition. These
distributions are very abstract ideas and while the web apps we use to
demonstrate their construction can help make sense of a single ``dot''
on the distribution, students commonly lose the forest for the trees.
\citet{wild2017} proposed the use of scaffolded animations to help
students hone their intuition about sampling/randomization variation and
to discover the utility of the bootstrap and permutation distributions.
While the animations discussed by \citet{wild2017} to visualize
randomization variation appear to be quite useful in communicating this
complex idea to students, a ``formal'' distribution of is not necessary
to introduce the core ideas behind hypothesis testing. Instead, we can
ask students to try to identify the data plot among a small set of decoy
plots generated by permutation resampling and link this simple
perceptual task with fundamental ideas of statistical inference.

In this article, we discuss how to use the lineup protocol from visual
inference to help students differentiate between different forms of
signal and noise and to better understand the nuances of statistical
significance. Section \ref{sec:vizinf} presents an overview of the
lineup protocol. Section \ref{sec:intro} presents examples of how the
lineup protocol can be used in the first course, and Section
\ref{sec:othercourses} presents additional examples of its use
throughout the curriculum. We conclude with a brief summary and
discussion in Section \ref{sec:conclusion}.

\section{Visual inference}
\label{sec:vizinf}

As outlined by \citet{Cobb-2007uo}, most introductory statistics books
teach that classical hypothesis tests consist of (i) formulating null
and alternative hypotheses, (ii) calculating a test statistic from the
observed data, (iii) comparing the test statistic to a reference (null)
distribution, and (iv) deriving a \(p\)-value on which a conclusion is
based. This is still true for the first course after adapting it based
on the new GAISE guidelines, regardless of whether a simulation-based
approach is used \citep[cf.~][]{Lock2017, tintle2015, introstats}. In
visual inference, the \emph{lineup protocol} provides a direct analog
for each step of a hypothesis test \citep{Buja-2009bd}.

\begin{enumerate}
\def\labelenumi{\arabic{enumi}.}
\item
  \textbf{Competing claims}: Similar to a traditional hypothesis test, a
  visual test begins by clearly stating the competing claims about the
  model/population parameters.
\item
  \textbf{Test statistic}: A plot displaying the raw data or fitted
  model (call the \emph{observed plot}) serves as the test statistic.
  This plot must be chosen to highlight features of the data that are
  relevant to the hypotheses. For example, a scatterplot is a natural
  choice to examine whether or not two quantitative variables are
  correlated.
\item
  \textbf{Reference (null) distribution}: \emph{Null plots} are
  generated consistently with the null hypothesis and the set of all
  null plots constitutes the \emph{reference} (or \emph{null})
  \emph{distribution}. To facilitate comparison of the observed plot to
  the null plots, the observed plot is randomly situated in the field of
  null plots, just as a suspect is randomly situated among decoys in a
  police lineup. This arrangement of plots is called a \emph{lineup}.
\item
  \textbf{Assessing evidence}: If the null hypothesis is true, then we
  expect the observed plot to be indistinguishable from the null plots.
  Consequently, if the observer is able to identify the observed plot in
  a lineup, then this provides evidence against the null hypothesis. If
  one wishes to calculate a visual p-value, then lineups need to be
  presented to a number of independent observers for evaluation. While
  this is possible, it is not a productive discussion in most
  introductory courses that don't explore probability theory.
\end{enumerate}

\hypertarget{example-comparing-groups}{%
\subsection{Example: Comparing groups}\label{example-comparing-groups}}

As a first example of visual inference via the lineup protocol, consider
the creative writing experiment discussed by \citet{ramsey2013}. The
experiment was designed to explore whether motivation type (intrinsic or
extrinsic) impacted creativity scores. To evaluate this, creative
writers were randomly assigned to a questionnaire where they ranked
reasons they write: one questionnaire listed intrinsic motivations and
the other listed extrinsic motivations. After completing the
questionnaire, all subjects wrote a Haiku about laughter, which was
graded for creativity by a panel of poets. \citet{ramsey2013} discuss
how to conduct a permutation test for the difference in mean creativity
scores between the two treatment groups. Below, we illustrate the steps
of a visual test.

\begin{enumerate}
\def\labelenumi{\arabic{enumi}.}
\item
  A visual test begins identically to a traditional hypothesis test by
  clearly stating the competing claims about the model/population
  parameters. In a first course, this could be written as:
  \(H_0: \mu_{\rm intrinsic} - \mu_{\rm extrinsic} = 0\)
  vs.~\(H_0: \mu_{\rm intrinsic} - \mu_{\rm extrinsic} \ne 0\).
\item
  In a visual test, plots take the role of test statistics
  \citep{Buja-2009bd}. In this situation, we must choose a plot (test
  statistic) that can highlight the difference in average creativity
  scores between the intrinsic and extrinsic treatment groups. Figure
  \ref{fig:data_plot} displays boxplots of creative writing scores by
  treatment group where a dot is used to represent the sample mean for
  each group, though other graphics could be used. There is an apparent
  difference in the distribution of the scores---the average score for
  the intrinsic group appears to be larger--but is it an important
  (i.e.~significant) difference?

  \begin{figure}
  \centering
  \includegraphics{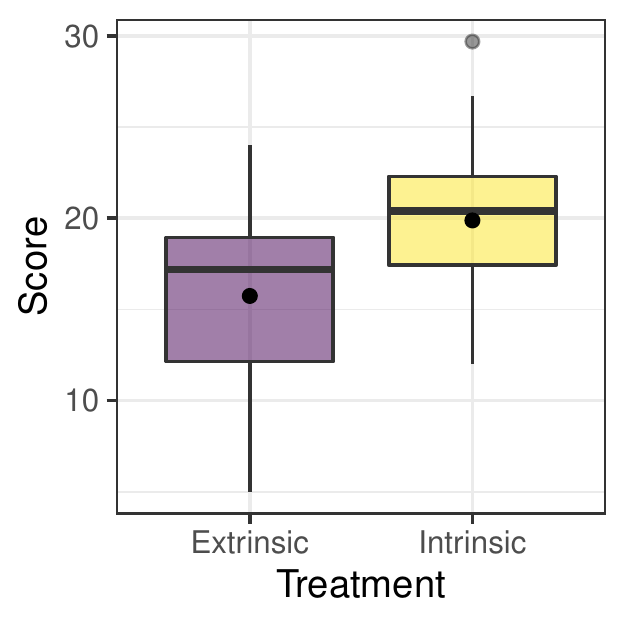}
  \caption{\label{fig:data_plot} Boxplots of the original creative
  writing scores by treatment group. The dot represents the mean of each
  group.}
  \end{figure}
\item
  To understand whether the observed (data) plot provides evidence of a
  significant difference, we must understand the behavior of our test
  statistic under the null hypothesis. To do this, we generate
  \emph{null plots} consistent with the null hypothesis and the set of
  all null plots constitutes the reference distribution. To facilitate
  comparison of the data plot to the null plots, the data plot is
  randomly situated in the field of null plots. This arrangement of
  plots is called a \emph{lineup}. Figure \ref{fig:lineup} shows one
  possible lineup for the creative writing experiment. The 19 null plots
  were generated via permutation resampling, and the data plot was
  randomly assigned to panel \#4.

  \begin{figure}
  \centering
  \includegraphics{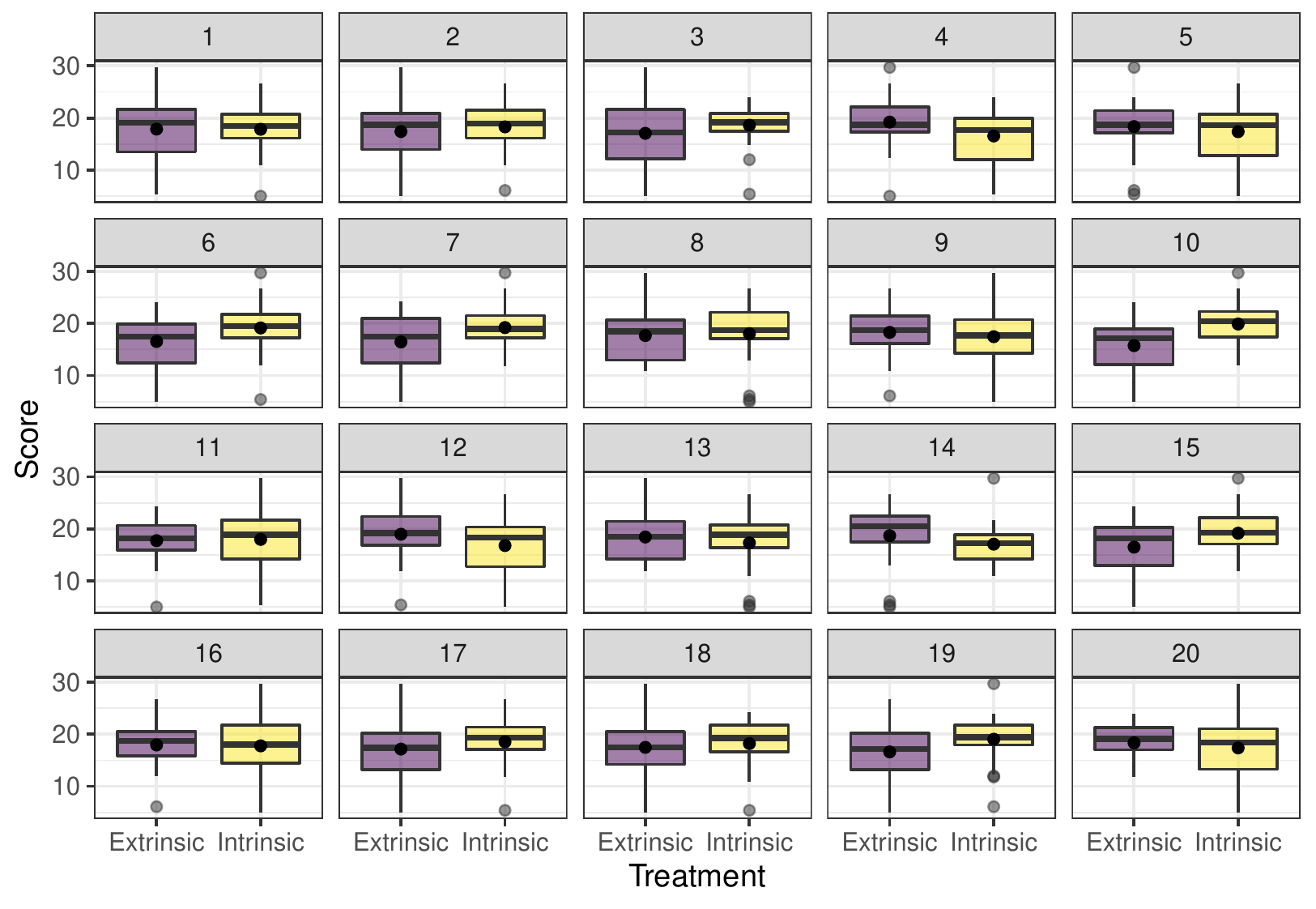}
  \caption{\label{fig:lineup} A lineup consisting of 19 null plots
  generated via permutation resampling and the original data plot for
  the creative writing study. The data plot was randomly placed in panel
  \#4.}
  \end{figure}
\item
  If the null hypothesis is true, then we expect the data plot to be
  indistinguishable from the null plots. Thus, if one is able to
  identify the data plot in panel \#4 of Figure \ref{fig:lineup}, then
  this provides evidence against the null hypothesis. We will not
  discuss the process of calculating a visual \(p\)-value, as the
  pedagogical value of the lineup protocol is in visualizing signal and
  noise.
\end{enumerate}

\section{Using visual inference in introductory statistics}
\label{sec:intro}

In this section, we discuss how to use the lineup protocol in the
introductory setting to introduce students to the logic of hypothesis
testing and to help students interpret new statistical graphics. The
goal is to provide examples of how this can be done, not to provide an
exhaustive list of possibilities.

\hypertarget{introducing-simulation-based-inference}{%
\subsection{Introducing (simulation-based)
inference}\label{introducing-simulation-based-inference}}

The strong parallels between visual inference and classical hypothesis
testing make it a natural way to introduce the idea of statistical
significance without getting bogged down in minutiae/controversy of
p-values, or the technical issues of describing a simulation procedure
before students understand why that is important. All students
understand the question ``which one of these plots is not like the
others,'' and this common understanding generates fruitful discussion
about the underlying inferential thought process without the need for a
slew of definitions. Below is an outline of a class activity discussing
the creative writing experiment to introduce the inferential thought
process.

\hypertarget{outline-of-activity}{%
\subsubsection{Outline of activity}\label{outline-of-activity}}

This activity is designed to be completed in groups of three or four
students. We have found that this group size allows all students to
contribute to the discussion, and is also conducive to assigned roles
\citep{Roseth2008-xx}, if you find that helps foster discussion in your
classroom.

\textbf{Competing claims.} To begin, we have students discuss what
competing claims are being investigated. We encourage them to write
these in words before linking them with the mathematical notation they
saw in the reading prior to class. The most common answer is: ``there is
no difference in the average creative writing scores for the two groups
vs.~there is a difference in the average creative writing scores for the
two groups.'' During the debrief, I make sure to link this to the
appropriate notation.

\textbf{EDA review.} Next, we have students discuss what plot types
would be most useful to investigate this claim. It's important to ask
students why they selected a specific plot type, as this reinforces
links to key ideas from exploratory data analysis.

\textbf{Lineup evaluation.} Most students recognize that side-by-side
boxplots, faceted histograms, or overlayed density plots are reasonable
choices to display the relevant aspects of the distribution of creative
writing scores for each group. We then provide a lineup of side-by-side
boxplots to evaluate (we do place a dot at the sample mean for each
group), such as the one shown in Figure \ref{fig:data_plot}. At this
point, we do not give the details behind the creation of null plots, we
simply tell students that one plot is the observed data while the other
nineteen agree with the null hypothesis. We ask students to (i) choose
which plot is the most different from the others and (ii) explain why
they chose that plot. Once each student has had time to make this
assessment (usually about one minute) we ask the groups to discuss and
defend their choices.

\textbf{Lineup discussion.} Once all of the groups have evaluated their
lineups and discussed their reasoning, we regroup for a class
discussion. During this discussion, we reveal which panel contains the
observed data (panel \#4 of Figure \ref{fig:lineup}), and display these
data on a slide so that we can point to particular features of the plot
as necessary. After revealing the observed data, we have students return
to their groups to discuss whether they chose the real data and whether
their choices support either of the competing claims. Once the class
regroups and thoughts are shared, we make sure that the class realizes
that an identification of the observed data provides evidence against
the null hypothesis (though I always hope students will be the ones
saying this).

\hypertarget{interpreting-unfamiliar-plots}{%
\subsection{Interpreting unfamiliar
plots}\label{interpreting-unfamiliar-plots}}

You can also utilize the lineup protocol in the first course to
introduce new and unfamiliar plot types. For example, we have found many
introductory students struggle to interpret residual plots. In this
situation, the lineup protocol helps students tune their understanding
of what constitutes an ``interesting'' pattern (i.e.~signal).

\hypertarget{residual-plots}{%
\subsubsection{Residual plots}\label{residual-plots}}

Interpreting residual plots is fraught with common errors, and we have
found that, regardless of our valiant attempts to explain what ``random
noise'' or ``random deviations from a model'' might look like, there is
no substitute for first hand experience. In this section, we outline a
class activity/discussion that we use to help train students to
interpret residual plots. This activity takes place after a brief
introduction to residual plots is given in class (or video if in a
flipped classroom). Again, we suggest that students complete such an
activity in small groups.

\begin{figure}

{\centering \includegraphics{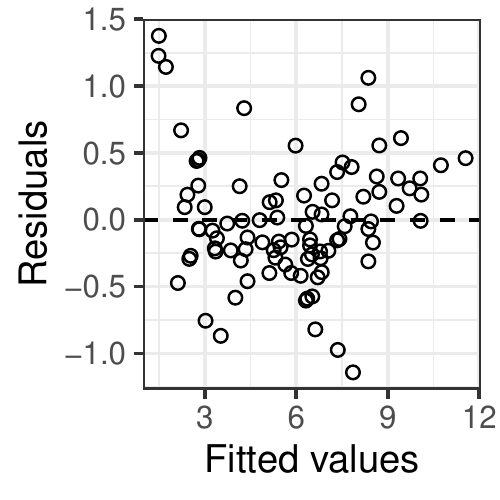} 

}

\caption{\label{fig:residplot} A residual plot for a simple linear regression model. Is there evidence that the model is insufficient?}\label{fig:residual plot}
\end{figure}

\textbf{Model fitting.} To begin, we have students fit a simple linear
regression model, write down what a residual is (in both words and using
notation), and then create a first residual plot, such as Figure
\ref{fig:residplot}.

\textbf{Interpreting residual plots.} Next, we pose the question: ``Does
this residual plot provide evidence of a model deficiency?'' This
provides students time to formalize their decision, and link it to
specific features of the residual plot upon which they based their
decision.

\begin{figure}
\centering
\includegraphics{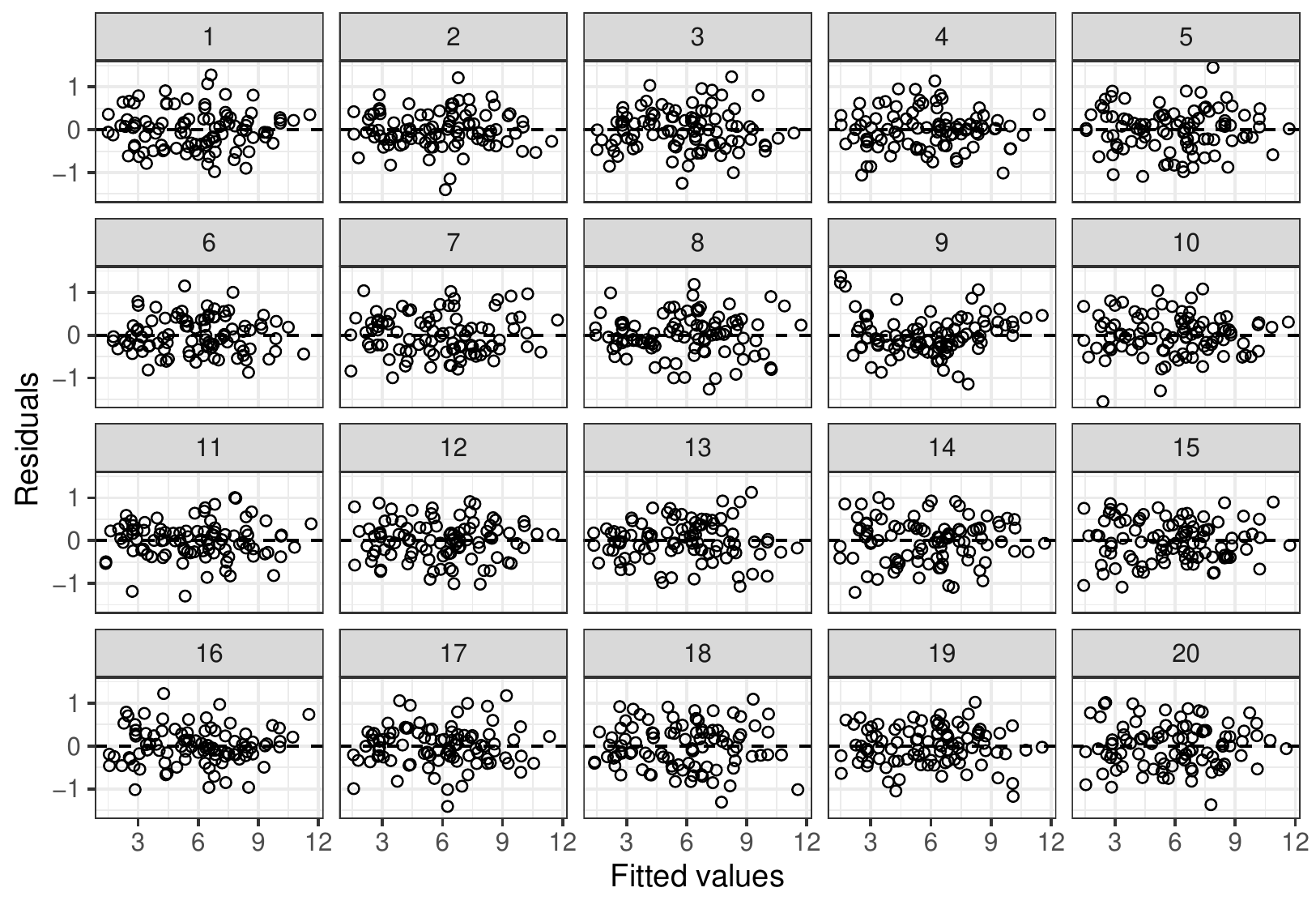}
\caption{\label{fig:lineupresid} A lineup of residual plots. The null
plots are generated via a parametric bootstrap from the fitted model.
The observed data are shown in panel \#9.}
\end{figure}

\textbf{Lineup evaluation.} Once students have carefully interpreted the
observed residual plot, we have them generate via a Shiny app (or
present them with) a lineup where their data plot has been randomly
situated in a field of null plots, such as the plot shown in Figure
\ref{fig:lineupresid}. Here, the null plots have been generated using
the parametric bootstrap, but the residual or non-parametric bootstraps
are other viable choices. We avoid the details of how the null plots
were generated, but this depends on the goals for your class. Once the
lineup has been generated, we ask students to (i) identify which panel
contains the observed residual plot, (ii) describe patterns they
observed in three null plots, and (iii) decide whether/how the observed
residual plot is systematically different from the null plots. We ask
students to answer the first question on their own before discussing
their choice with their group, and typically give one or two minutes of
individual ``think time'' prior to group discussion.

\textbf{Debrief.} Once all of the groups have evaluated their lineups
and discussed their reasoning, it is important to regroup for a class
discussion. This allows you to reveal the observed residual plot and
revisit key points about residual plots and their interpretation.

\textbf{Teaching tips}

\begin{itemize}
\item
  In Figure \ref{fig:lineupresid}, the observed residual plot in panel
  \#9 is systematically different from the null plots. While this is one
  example we use in class, we also recommend a parallel example where
  there is no discrepancy between the data and the model.
\item
  Depending on your course goals, follow-up discussions about the design
  of residual plots could be injected to the end of this activity. For
  example, you could provide students with a second version of the
  lineup where LOESS smoothers have been added to each panel and ask
  students what features of the residual plot the smoother highlights.
\item
  An alternative activity first has students use the
  \emph{Rorschach protocol} \citep{Buja-2009bd} to look through a series
  of null plots, describing what they see, and then looking at a single
  residual plot.
\end{itemize}

\hypertarget{other-plot-types}{%
\subsubsection{Other plot types}\label{other-plot-types}}

Similar activities can be designed to introduce other statistical
graphics. Specifically, we have also found that lineups help students
learn to read normal quantile-quantile and mosaic plots (or stacked bar
charts).

\section{Using visual inference in other courses}
\label{sec:othercourses}

The utility of visual inference is not limited to introductory courses.
Whenever a new model is encountered intuition about diagnostic plots
must be rebuilt, and the lineup protocol helps students build this
intuition. As an example, consider diagnostics for binary logistic
regression models, a common topic in a second course.

\hypertarget{diagnostics-for-binary-logistic-regression}{%
\subsection{Diagnostics for binary logistic
regression}\label{diagnostics-for-binary-logistic-regression}}

\begin{figure}
\centering
\includegraphics{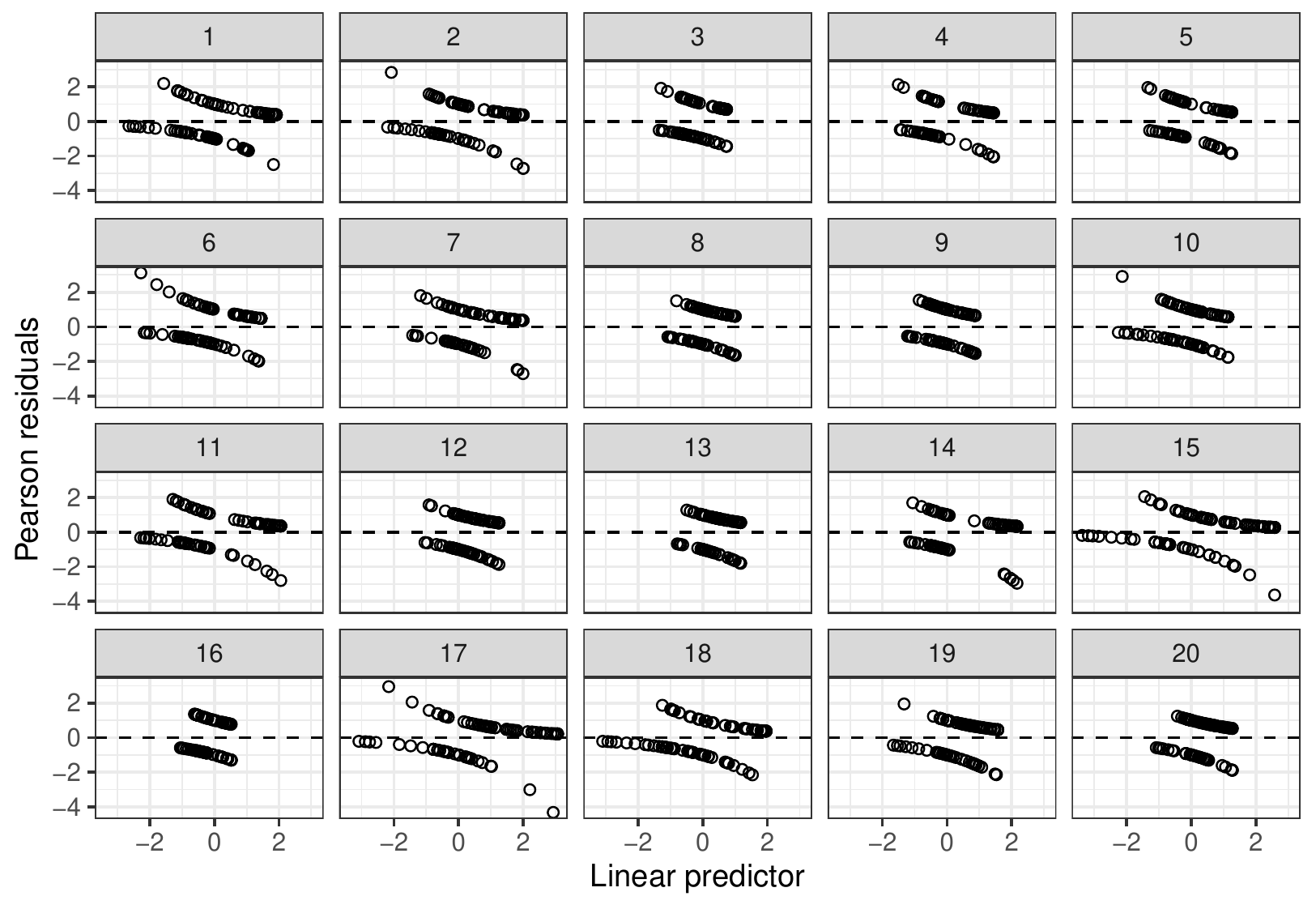}
\caption{\label{fig:logisticissue} A lineup for a deficient logistic
regression model. The data plot are simulated from a model with a
quadratic effect, while the null plots are simulated from a model with
only a linear effect. Can you identify the deficient plot?}
\end{figure}

Interpreting residual plots from binary logistic regression is
difficult, as plots of the residuals against the fitted values or
predictors often look similar for adequate and inadequate models. The
lineup protocol provides a framework for this discussion. For example,
you can simulate data from a model where a quadratic effect is needed,
but fit the data to a model with only a linear effect and extract the
Pearson residuals. Then, you can simulate the null plots from the model
with only the linear effect and extract the Pearson residuals. Figure
\ref{fig:logisticissue} shows a lineup created in this way. Having a
discussion surrounding this lineup in class will help pinpoint the
difficulty using conventional residual plots for model diagnoses.

After establishing the pitfalls of ``conventional'' residual plots for
binary logistic regression, you can introduce alternative strategies
(i.e., new diagnostic plots) and again use the lineup protocol to
calibrate student intuition. Below are two such examples.

\hypertarget{binned-residual-plots}{%
\subsubsection{Binned residual plots}\label{binned-residual-plots}}

\begin{figure}
\centering
\includegraphics{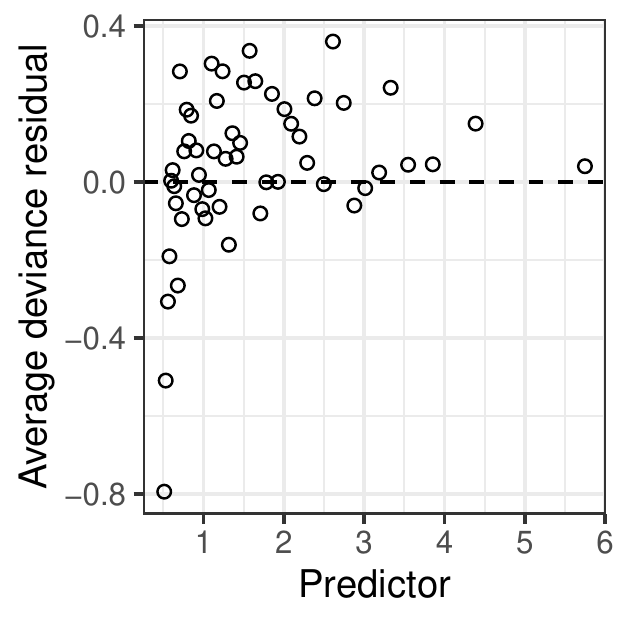}
\caption{\label{fig:binned} A binned residual plot from a simple binary
logistic regression model. The average deviance residual is plotted on
the \(y\)-axis for each of 54 bins on the \(x\)-axis.}
\end{figure}

\begin{figure}
\centering
\includegraphics{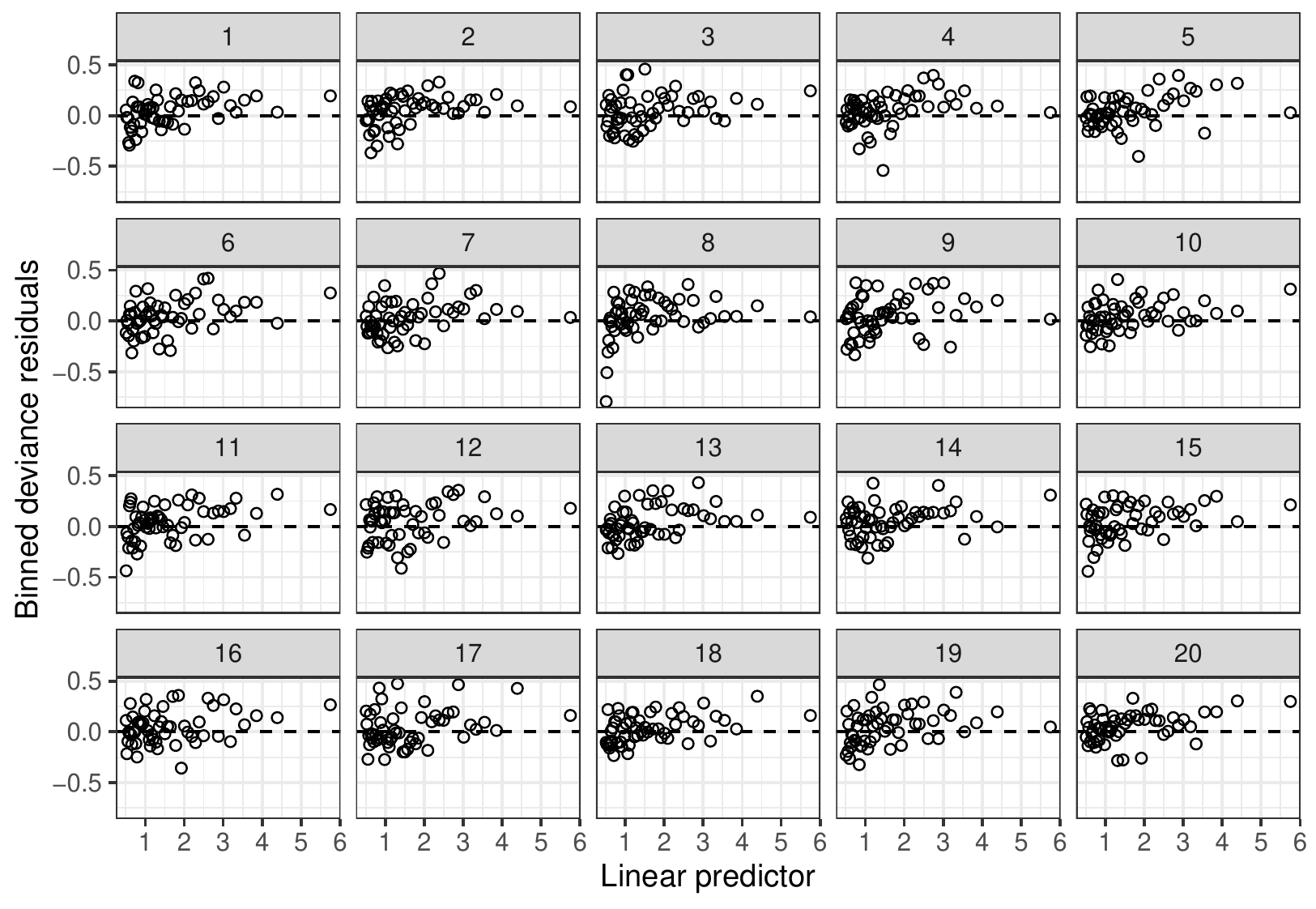}
\caption{\label{fig:binnedlineup} A lineup of binned residual plots from
a simple binary logistic regression model. The observed residuals are
shown in panel \#8 and clearly stand out from the field of null plots,
indicating a problem with linearity.}
\end{figure}

\citet{GelmanHill:2007} recommend using \emph{binned residual plots} to
explore possible violations of linearity for binary logistic regression.
A binned residual plot is created by calculating the average residual
value with bins that partition the \(x\)-axis. Figure \ref{fig:binned}
shows a binned residual plot from a simple binary logistic regression
model. The average deviance residual is plotted on the \(y\)-axis for
each of 54 bins on the \(x\)-axis. The number of bins is set to
\(\lfloor \sqrt{n} \rfloor\), but can be adjusted as with a histogram.
\citet{GelmanHill:2007} claim that these plots should behave much like
the familiar standardized residual plots from regression. If this is the
case, then Figure \ref{fig:binned} is indicative of nonlinearity.
However, rather than simply citing \citet{GelmanHill:2007} to students,
a lineup empowers them to investigate the behavior of this new plot
type. A lineup for these residuals is given in Figure
\ref{fig:binnedlineup}. As suspected, the data plot (panel \#8) stands
out from the field of null plots, indicating a deficiency with the
model. This investigation via a lineup can be framed as a whole class
discussion or as a group activity, similar to the activities already
outlined.

\hypertarget{empirical-logit-plots}{%
\subsubsection{Empirical logit plots}\label{empirical-logit-plots}}

A more-common alternative to the binned residual plot is the
\emph{empirical logit plot} \citep[c.f.,][]{stat2, ramsey2013}. An
empirical logit plot can be constructed for each explanatory variable by
calculating the adjusted proportion of ``successes'' within each
``group'' as \[
\widehat{p}_{\rm adj} = \frac{\text{number successes} + 0.5}{\text{number of cases} + 1},
\] and plotting
\(\log\left(\widehat{p}_{\rm adj} / (1- \widehat{p}_{\rm adj}) \right)\)
against the average value of a quantitative explanatory variable, or the
level of a categorical explanatory variable. For quantitative variables,
it is common to form groups by forming bins of roughly equal size.

\begin{figure}

{\centering \includegraphics[width=0.85\linewidth]{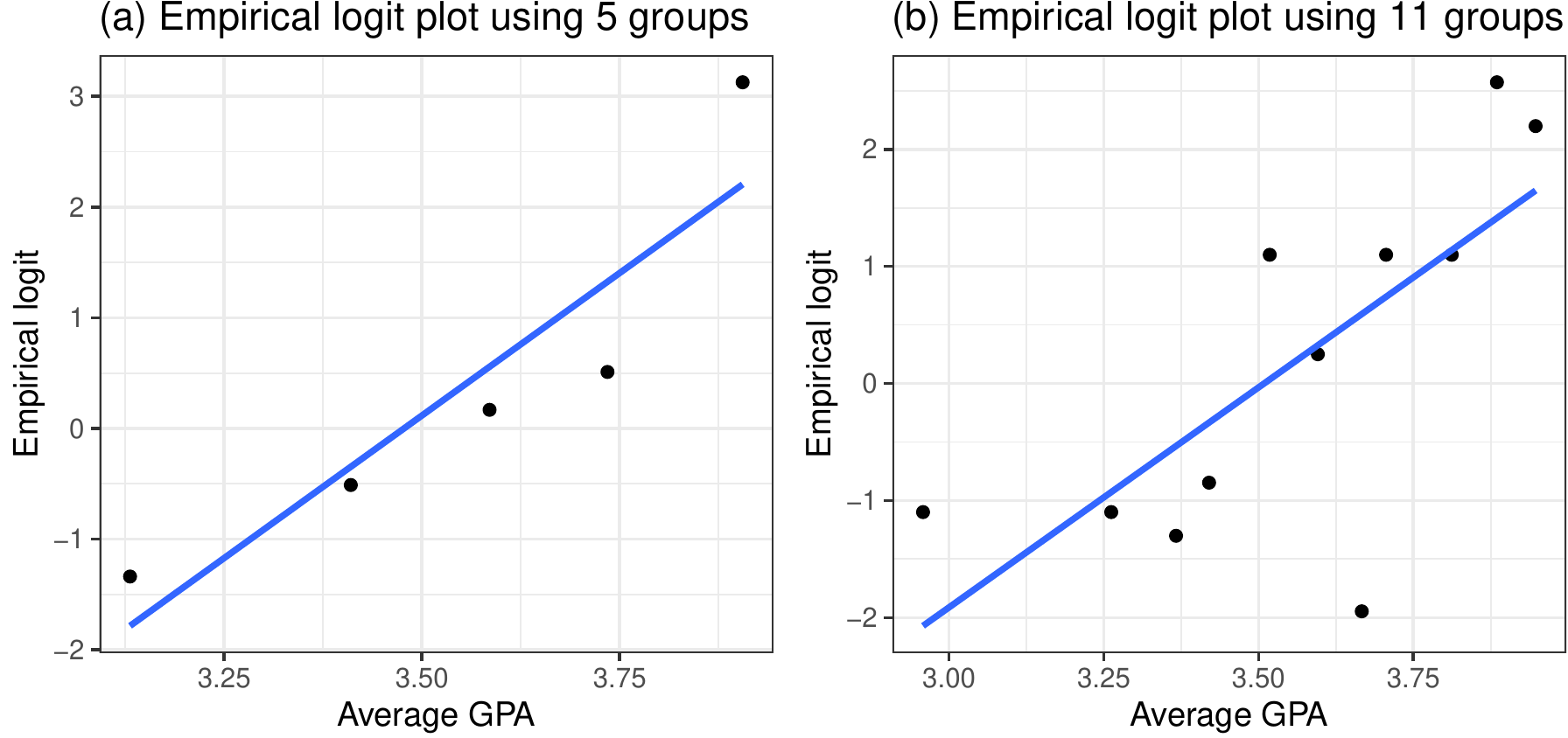} 

}

\caption{\label{fig:emplogitexample} Two empirical logit plots rendered for the same data set with $n=55$ observations. Panel (a) is rendered using 5 groups while panel (b) is rendered using 11 groups. The appearance of the plots changes substantially, often leading to confusion in intrepretation.}\label{fig:unnamed-chunk-7}
\end{figure}

While an empirical logit plot is quite straightforward to create, it can
be hard to interpret for smaller data sets where few groups are formed.
For example, \citet{stat2} use empirical logit plots to explore a binary
logistic regression model for medical school admission decisions based
on an applicant's average grade point average and render empirical logit
plots based on both 5 and 11 bins. Figure \ref{fig:emplogitexample}
shows recreations of these plots. Experimenting with the bin width
reveals the difficulty students may encounter determining whether
linearity is reasonable: the plot can change substantially based on the
binwidth. In our experience, students often see some indication of
non-linearity in the plot with 5 bins (Figure \ref{fig:emplogitexample}
(a)), whereas they think the plot with 11 bins (Figure
\ref{fig:emplogitexample} (b)) is reasonably linear.

\begin{figure}
\centering
\includegraphics{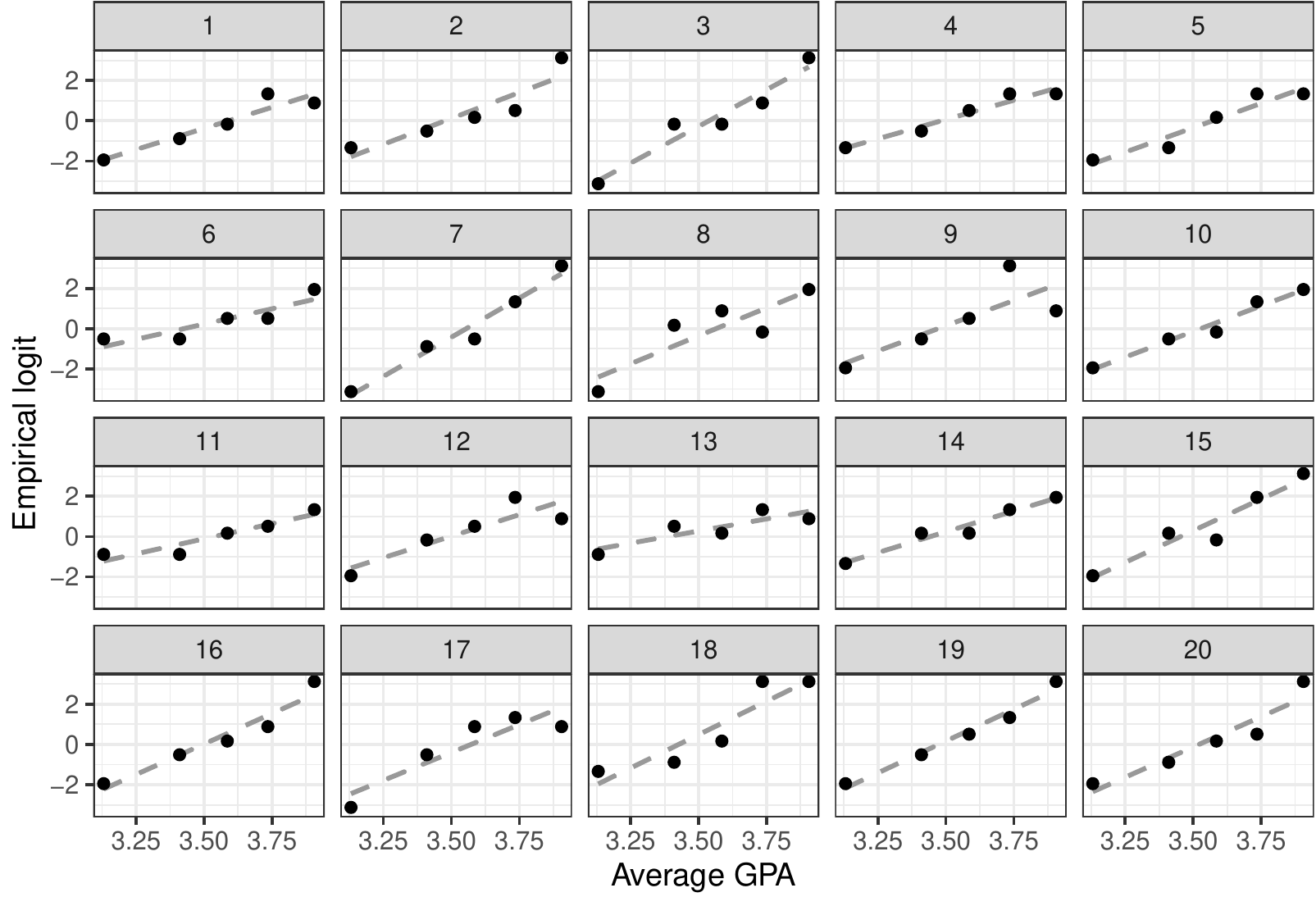}
\caption{\label{fig:emplogitlineup} A lineup of empirical logit plots
from a simple binary logistic regression model. The observed plot is
shown in panel \#2 and does not stand out from the field of null plots,
indicating no problem with linearity.}
\end{figure}

To help students interpret whether observed patterns on empirical logit
plots are problematic, we again appeal to the lineup protocol to enforce
a comparison between the observed plot and what is expected under the
model. Figure \ref{fig:emplogitlineup} displays a lineup of the
empirical logit plot created using 5 groups. The observed plot (panel
\#2) is difficult to pick out from the field of null plots, providing no
evidence that against linearity.

\hypertarget{diagnosing-other-models}{%
\subsection{Diagnosing other models}\label{diagnosing-other-models}}

In this section, we focused on using the lineup protocol to help
diagnose logistic regression models, but the approach is generally
applicable. If you have a plot highlighting some feature(s) of the
fitted model, then after simulating data from a ``correct'' model
(i.e.~one without model deficiencies), you can create a lineup to
interrogate the model. For example, \cite{Loy2017-fo} discuss how visual
inference can be used to diagnose multilevel models.

\hypertarget{implementation}{%
\section{Implementation}\label{implementation}}

\label{sec:implement}

All of the lineups presented in this paper were rendered in R \citep{r}.
A tutorial outlining this process using the ggplot2 \citep{ggplot2} and
nullabor \citep{nullabor} R packages is provided in the supplementary
materials. These tools allow you to customize lineups for class use, but
we do not recommend having introductory students grapple with this code.
For introductory students, we recommend providing handouts or slides
with pre-rendered lineups during class activities. Alternatively, we
have created a suite of Shiny apps \citep{shiny} where students can
upload data sets and render lineups. The current suite includes apps to
generate lineups to explore associations between groups, normal Q-Q
plots, and residual plots for simple linear regression models. Links to
the shiny apps can be found at
\url{https://aloy.rbind.io/project/classroom-viz-inf/}.

While we have found in-class activities where students use the lineup
protocol to explore new plots types and the inferential thought process
to be useful, these could also be assigned as homework problems or
pre-class exercises. Regardless of the venue, scaffolding the activity
to guide students and foster discussion/reflection is key. In the above
examples, we illustrated the approach that has worked well for our
students, but each instructor should adjust this to their own class,
teaching style, and student population. In addition, our discussion has
not been exhaustive, so we encourage instructors to identify additional
places in their curriculum where the lineup protocol can help their
students build intuition.

\hypertarget{conclusion}{%
\section{Conclusion}\label{conclusion}}

\label{sec:conclusion}

The lineup protocol provides a framework to help students learn to
interpret new statistical graphics and hone their intuition about what
constitutes an interesting feature/pattern. This is achieved by randomly
embedding the observed data plot into a field of decoy (null) plots.
Lineups provide a natural way to introduce new statistical graphics
throughout the statistics curriculum. At the introductory level, lineups
can help students learn to detect association in side-by-side boxplots
and mosaic plots, and detect problematic patterns in residual plots for
regression models. In more advanced courses, lineups can be used to
frame conversations about why conventional residual plots are
problematic for certain models and can improve a student's diagnostic
ability as they investigate new models.

The shift to permutation tests in introductory courses has lowered the
initial technical barriers to hypothesis testing; however, it still
requires an explanation of \emph{why} we need to resample and \emph{how}
we resample. Exploring lineups that you provide and making the analogy
to the police lineup (or alternatively the Sesame Street question:
``which one of these is not like the others'') introduces students to
the the logic behind testing without the need for these technical
discussions. This allows initial focus to be on the core concepts of
hypothesis testing rather than simultaneous focus on the core concepts
and the technical details. We have found that a wide range of students
understand why an inferential process is needed and what the findings
imply at a more intuitive level after grappling with questions such as
``which one of these plots is not like the others?'', ``how do you
know?'', and ``what does this mean about your initial claim?'' In
addition, permutation tests logically follow the lineup protocol,
providing students with the details behind the generation of the
null/decoy plots and ways to formalize the strength of evidence against
an initial claim.

Finally, the lineup protocol equips students with a rigorous tool for
visual investigation that is applicable outside of the classroom. This
not only prepares students to explore unfamiliar models or graphics in
their own statistical analyses, but can facilitate ``teaser''
conversations about advanced models for majors. For example, if you
introduce your students to the lineup protocol in a modeling course,
then you can show a lineup of choropleth maps and discuss spatial
statistics as a potential area of future study.

\section*{Supplementary materials}

\begin{itemize}
\item
  A tutorial on using nullabor and ggplot2 to create lineups for common
  topics in introductory statistics can be found at
  \url{https://aloy.github.io/classroom-vizinf/}.
\item
  A suite of Shiny apps that creates lineups for common topics in
  introductory statistics can be found at
  \url{https://aloy.rbind.io/project/classroom-viz-inf/}
\end{itemize}

\section*{Acknowledgements}

The author wishes to thank the editorial board the StatTLC blog for
their thoughts on an early version of this manuscript.

\bibliographystyle{agsm}
\bibliography{classroomVizInf.bib}

@ARTICLE{wild2017,
  title   = "Accessible Conceptions of Statistical Inference: Pulling Ourselves
             Up by the Bootstraps",
  author  = {Wild, Chris J. and Pfannkuch, Maxine  and Regan, Matt  and Parsonage, Ross},
  journal = "International Statistical Review",
  volume  =  85,
  number  =  1,
  pages   = "84--107",
  year    =  2017,
  url     = "https://onlinelibrary.wiley.com/doi/full/10.1111/insr.12117",
  doi     = "10.1111/insr.12117"
}

@MISC{gaise2016,
     title = {Guidelines for Assessment and Instruction in
Statistics Education College Report 2016},
     author = {{GAISE College Report ASA Revision Committee}},
     group = {GAISE},
     year = {2016},
     institution = {University of Zurich, Department of Informatics},
     url = { http://www.amstat.org/education/gaise}
}

@book{ramsey2013,
  title={The Statistical Sleuth: A course in Methods of Data Analysis},
  author={Ramsey, Fred and Schafer, Daniel},
  year={2013},
  edition = {3},
  address = {Boston, MA},
  publisher={Cengage Learning}
}

@ARTICLE{Loy2017-fo,
  title     = "Model Choice and Diagnostics for Linear {Mixed-Effects} Models
               Using Statistics on Street Corners",
  author    = "Loy, Adam and Hofmann, Heike and Cook, Dianne",
  journal   = "Journal of Computational and Graphical Statistics",
  volume    =  26,
  number    =  3,
  pages     = "478--492",
  month     =  jul,
  year      =  2017,
  url       = "http://dx.doi.org/10.1080/10618600.2017.1330207",
  doi       = "10.1080/10618600.2017.1330207"
}

@book{agresti2017,
  title={Statistics: The Art and Science of Learning from Data},
  author={Alan Agresti and Christine A. Franklin and Bernhard Klingenberg},
  year={2017},
  edition={4th},
  publisher={Prentice Hall}
}

@book{tintle2015,
  title={Introduction to statistical investigations},
  author={Tintle, Nathan and Chance, Beth L and Cobb, George W and Rossman, Allan J and Roy, Soma and Swanson, Todd and VanderStoep, Jill},
  year={2015},
  publisher={John Wiley \& Sons},
  address = {Danvers, MA}
}

@book{introstats,
  title={Intro Stats},
  author={De Veaux, Richard and Velleman, Paul and Bock, David},
  year={2018},
  edition = {5},
  publisher = {Pearson},
  address = {Boston, MA}
}

@book{chihara2011,
  title={Mathematical Statistics with Resampling and R},
  author={Chihara, Laura and Hesterberg, Tim},
  year={2011},
  publisher={Wiley}
}

@ARTICLE{Tintle2015-yv,
  title     = "Combating {Anti-Statistical} Thinking Using {Simulation-Based}
               Methods Throughout the Undergraduate Curriculum",
  author    = "Tintle, Nathan and Chance, Beth and Cobb, George and Roy, Soma
               and Swanson, Todd and VanderStoep, Jill",
  journal   = "The American Statistician",
  publisher = "Taylor \& Francis",
  volume    =  69,
  number    =  4,
  pages     = "362--370",
  month     =  oct,
  year      =  2015
}

@ARTICLE{Cobb2011-vz,
  title   = "Teaching statistics: Some important tensions",
  author  = "Cobb, George W",
  journal = "Chil. J. Stat.",
  volume  =  2,
  number  =  1,
  pages   = "31--62",
  year    =  2011
}

@INPROCEEDINGS{Tintle2014-vt,
  title     = "Quantitative Evidence for the Use of Simulation and
               Randomization in the Introductory Statistics Course",
  booktitle   = "Sustainability in statistics education. Proceedings of the
                 Ninth International Conference on Teaching Statistics
                 ({ICOTS9}, July, 2014), Flagstaff, Arizona, {USA}.",
  author    = "Tintle, Nathan and Rogers, Ally and Chance, Beth
               and Cobb, George and Rossman, Allan and Roy, Soma and
               Swanson, Todd and VanderStoep, Jill",
  editor    = { Makar, K and de Sousa, B. and Gould, R.},             
  institution = "International Statistical Institute",
  year      =  2014
  }

@article{Buja-2009bd,
author = {Buja, A and Cook, D and Hofmann, H and Lawrence, M and Lee, E K and Swayne, D F and Wickham, H},
title = {Statistical inference for exploratory data analysis and model diagnostics},
journal = {Philosophical Transactions of the Royal Society A},
year = {2009},
volume = {367},
number = {1906},
pages = {4361--4383},
month = oct
}

@article{Cobb-2007uo,
author = {Cobb, G W},
title = {The introductory statistics course: a Ptolemaic curriculum?},
journal = {Technology Innovations in Statistics Education},
year = {2007},
volume = {1},
issue = {1}
}

@book{stat2,
author = {Ann Cannon and George W. Cobb; and Bradley A. Hartlaub and Julie M. Legler and 
          Robin H. Lock and Thomas L. Moore and Allan J. Rossman and Jeffrey A. Witmer},
title = {{STAT2}: Modeling with Regression and {ANOVA}},
publisher = {MacMillan},
year = {2018},
edition = {2nd}
}

@book{Lock2017,
author = {Lock, Robin and Frazer Lock, Patti and Lock Morgan, Kari and Lock, Eric and Lock, Dennis},
title = {Statistics: Unlocking the Power of Data},
publisher = {John Wiley \& Sons},
year = {2017},
edition = {2nd},
address = {Hoboken}
}

@article{Maurer-2014te,
author = {Maurer, Karsten and Lock, Dennis},
title = {Comparison of Learning Outcomes for Randomization-Based and Traditional Inference Curricula in a Designed Educational Experiment},
year = {2014},
pages = {1--18},
}

@Manual{nullabor,
    title = {{nullabor}: Tools for Graphical Inference},
    author = {Hadley Wickham and Niladri Roy Chowdhury and Di Cook},
    year = {2014},
    note = {{R} package version 0.3.1},
    url = {http://CRAN.R-project.org/package=nullabor},
}

@article{Tintle-2011vo,
author = {Tintle, N and VanderStoep, J and Holmes, V L},
title = {Development and assessment of a preliminary randomization-based introductory statistics curriculum},
journal = {Journal of Statistics Education},
volume = {19},
issue = {1},
year = {2011}
}

@article{Tintle-2012td,
author = {Tintle, N L and Topliff, K and VanderStoep, J},
title = {Retention of statistical concepts in a preliminary randomization-based introductory statistics curriculum},
journal = {Statistics Education Research Journal},
year = {2012},
volume = {11},
issue = {1},
pages = {21--40}
}

@book{GelmanHill:2007,
  address = {New York},
  author = {Gelman, Andrew and Hill, Jennifer},
  publisher = {Cambridge University Press},
  title = {Data analysis using regression and multilevel/hierarchical models},
  year = 2007
}

@ARTICLE{Hildreth2018,
  title   = "Comparing Student Success and Understanding in Introductory
             Statistics under Consensus and {Simulation-Based} Curricula",
  author  = "Hildreth, Laura A and Robison-Cox, Jim and Schmidt, Jade",
  journal = "Statistics Education Research Journal",
  volume  =  17,
  number  =  1,
  pages   = "103--120",
  year    =  2018
  }

@Book{ggplot2,
    author = {Hadley Wickham},
    title = {ggplot2: Elegant Graphics for Data Analysis},
    publisher = {Springer-Verlag New York},
    year = {2016},
    isbn = {978-3-319-24277-4},
    url = {https://ggplot2.tidyverse.org},
  }

@Manual{ggformula,
    title = {ggformula: Formula Interface to the Grammar of Graphics},
    author = {Daniel Kaplan and Randall Pruim},
    year = {2019},
    note = {R package version 0.9.1},
    url = {https://CRAN.R-project.org/package=ggformula},
  }

@ARTICLE{Pruim2017-uc,
  title   = "The mosaic Package: Helping Students to 'Think with Data' Using
             {R}",
  author  = "Pruim, Randall and Kaplan, Daniel T and Horton, Nicholas J",
  journal = "R Journal",
  volume  =  9,
  number  =  1,
  pages   = "77--102",
  year    =  2017,
  url     = "https://journal.r-project.org/archive/2017/RJ-2017-024/RJ-2017-024.pdf",
  issn    = "2073-4859"
}

@Manual{infer,
    title = {{infer}: Tidy Statistical Inference},
    author = {Andrew Bray and Chester Ismay and Evgeni Chasnovski and Ben Baumer and Mine Cetinkaya-Rundel},
    year = {2019},
    note = {R package version 0.5.1},
    url = {https://CRAN.R-project.org/package=infer},
  }

@Manual{r,
    title = {R: A Language and Environment for Statistical Computing},
    author = {{R Core Team}},
    organization = {R Foundation for Statistical Computing},
    address = {Vienna, Austria},
    year = {2019},
    url = {https://www.R-project.org/},
  }

@ARTICLE{Roseth2008-xx,
  title     = "Collaboration in Learning and Teaching Statistics",
  author    = "Roseth, Cary J and Garfield, Joan B and Ben-Zvi, Dani",
  journal   = "Journal of statistics education: an international journal on the
               teaching and learning of statistics",
  volume    =  16,
  number    =  1,
  year      =  2008,
  url       = "https://doi.org/10.1080/10691898.2008.11889557",
  doi       = "10.1080/10691898.2008.11889557"
}

@Manual{shiny,
    title = {shiny: Web Application Framework for R},
    author = {Winston Chang and Joe Cheng and JJ Allaire and Yihui Xie and Jonathan McPherson},
    year = {2019},
    note = {R package version 1.4.0},
    url = {https://CRAN.R-project.org/package=shiny},
  }

\end{document}